%% file: JURSE2019_template_nocopyright.tex
\documentclass[conference]{IEEEtran}
\input{preambol}

\IEEEoverridecommandlockouts
\usepackage{cite}
\usepackage{amsmath,amssymb,amsfonts}
\usepackage{algorithmic}
\usepackage{graphicx}
\usepackage{textcomp}
\usepackage{xcolor}
\def\BibTeX{{\rm B\kern-.05em{\sc i\kern-.025em b}\kern-.08em
    T\kern-.1667em\lower.7ex\hbox{E}\kern-.125emX}}

\begin{document}
\title{A Novel Cost Function for Despeckling using Convolutional Neural Networks\\}


\author{
	\IEEEauthorblockN{ Giampaolo Ferraioli}
	\IEEEauthorblockA{\textit{Dipartimento di Scienze e Tecnologie} \\
		\textit{Universit\`{a} di Napoli Parthenope}\\
		Napoli,Italy \\
		giampaolo.ferraioli@uniparthenope.it}
	\and
	\IEEEauthorblockN{ Vito Pascazio}
	\IEEEauthorblockA{\textit{Dipartimento di Ingegneria} \\
		\textit{Universit\`{a} di Napoli Parthenope}\\
		Napoli,Italy \\
		vito.pascazio@uniparthenope.it}
	\and
	\IEEEauthorblockN{ Sergio Vitale}
	\IEEEauthorblockA{\textit{Dipartimento di Ingegneria} \\
		\textit{Universit\`{a} di Napoli Parthenope}\\
		Napoli,Italy \\
		sergio.vitale@uniparthenope.it}
}

\maketitle



\begin{abstract}
	Removing speckle noise from SAR images is still an open issue. It is well know that the interpretation of SAR images is very challenging and despeckling algorithms are necessary to improve the ability of extracting information. An urban environment makes this task more heavy due to different structures and to different objects scale. Following the recent spread of deep learning methods related to several remote sensing applications, in this work a convolutional neural networks based algorithm for despeckling is proposed. The network is trained on simulated SAR data. The paper is mainly focused on the implementation of a cost function that takes account of both spatial consistency of image and statistical properties of noise. 
	\\
\end{abstract}


\section{Introduction}
\label{sec:intro}
\begin{figure}
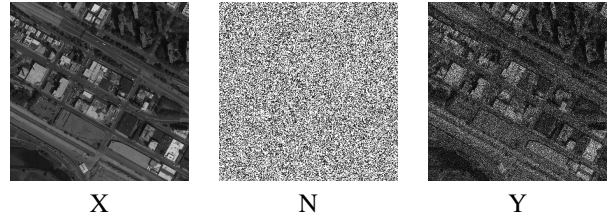

	\begin{tabular}{ccc}
		\image{ref} & \image{speckle} &\image{noisy}\\
		X	& N & Y \\
	\end{tabular}
	\caption{Simulated SAR image in hypothesis of multiplicative speckle}
	\label{fig: speckle}
\end{figure}

In the last decades, remote sensing has continuously grown providing more and more images of the planet. The way to extract useful informations is still an open issue,  even more when we are dealing with SAR sensors. SAR images are affected by multiplicative noise called speckle, that impairs performances of different tasks such as classification, object detection and segmentation. In fact, in these years a very big area of research has grown to tackle this problem and a lot of despeckling algorithms have been proposed. As said before, speckle is a multiplicative noise given by the interaction of electromagnetic fields scattered in different directions from a rough surface. Let's consider $Y$ a SAR image, it can be expressed as  \cite{argenti2013tutorial}:
\\
\begin{equation}
Y = f(X,N) = X\cdot N
\label{eq: sar formation}
\end{equation}
\\
where $X$ is the noise-free image and $N$ is the multiplicative speckle. In the hypothesis of fully developped speckle, its distribution is known and, for an intensity image, it is a Gamma distribution \cite{touzi2002}:
\\
\begin{equation}
p(N) = \frac{1}{\Gamma(L)} N^L e^{-NL}
\label{eq: speckle distribution}
\end{equation}
\\
where $L$ is the number of looks of SAR image, (Fig. \ref{fig: speckle}).
An ideal despeckling filter will remove the noise without introducing artefacts and preserving the spatial informations.
The despeckling filters are usually divided in two categories: local and non local filters. The formers as Lee\cite{Lee1980}, Enhanced Lee\cite{Lopes1990} and Kuan filter\cite{Kuan1985} rely on similarity between the target and its adjacent pixels. The latter as Patch Probabilist Based (PPB)\cite{Deledalle2009}, SAR-BM3D\cite{Parrilli2012}, NL-SAR \cite{Deledalle2015} look for similarity in a wider window search.
Nowadays, with the increasing of deep learning solutions in a lot of fields related to image processing, another branch of filters has born. Indeed, in the last years also convolutional neural networks (CNN) based solutions have been proposed such as \cite{Wang2017}, \cite{Chierchia2017}.
Using CNNs for despeckling is quite challenging because the lack of a clean reference: once a real SAR image is acquired, there is no possibility to have a speckle free image to use as reference.

The trends to overcome this problem are mainly two:
\begin{itemize}
	\item training a network to perform one of despeckling filter as in \cite{Chierchia2017}, in which a CNN is proposed to perform multilook when there is no chance to have several acquisitions of same data;
	\item training on simulated data as in \cite{Wang2017}.
\end{itemize}  
As in \cite{Wang2017}, in this work SAR simulated data are used. Clean images are taken from three datasets: UCID, BSD\cite{MartinFTM01} and scraped Google Maps\cite{Isola16}. The Google Maps dataset is composed by images in urban environment, instead in UCID and BSD there are generic images.

\begin{figure*}
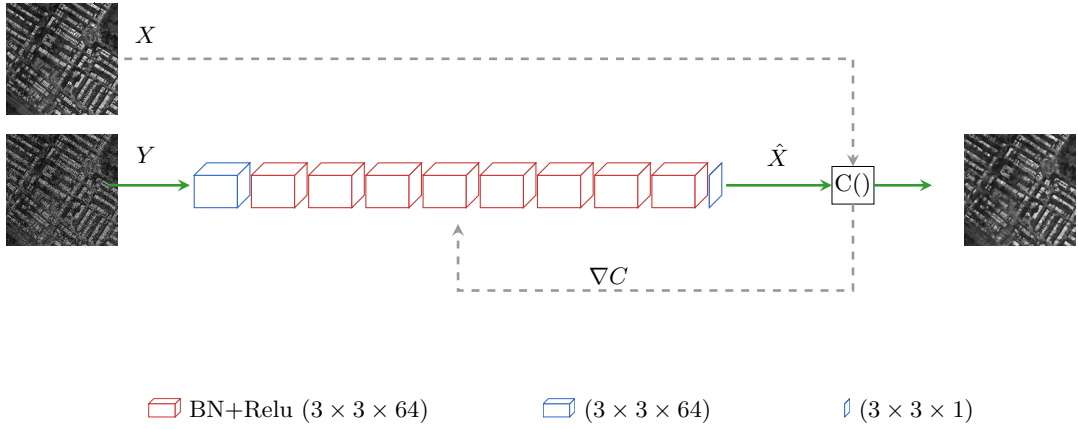

	\centering
	\image{net}
	\caption{Top-level workflow of the despeckling CNN.}
	\label{fig: Net}
\end{figure*}

\section{Proposed Approach}
In this work a deep learning solution for despeckling is proposed. It is focused on the use of deep convolutional neural networks and on their ability to predict the noise and provide a filtered image in which spatial and statistical details are preserved.
\subsection{Convolutional Neural Networks}
A CNN is composed by a combination of several layers, connected in different ways (cascade, parallel, loop). Each layer can perform different function: convolution, pooling, non-linearities. 

A generic layer provides a set of $ M $ so-called {\em feature maps}. Higher is the level of the layer, more abstract is its output and more representative of overall interaction  between layers.
So the $l$-th generic convolutional layer,  for $N$-bands input $\bx^{(l)}$, yields an $M$-band 
output $\bz^{(l)}$ 
\setlength{\abovedisplayskip}{10pt}
\setlength{\belowdisplayskip}{10pt}

\[
\bz^{(l)} = \bw^{(l)} \ast \bx^{(l)} + \bbb^{(l)},
\]
\\
whose $m$-th component is a combination of 2D convolutions:

\[
\bz^{(l)}(m,\cdot,\cdot) = \sum_{n=1}^N  \bw^{(l)}(m,n,\cdot,\cdot) \ast \by^{(l)}(n,\cdot,\cdot)+ \bbb^{(l)}(m).
\]
\setlength{\belowdisplayskip}{10pt}

The tensor $\bw$ is a set of $M$ convolutional $N\times(K\times K)$ kernels, with a $K\times K$ spatial support (receptive field),
while $\bbb$ is a $M$-vector bias.
These parameters, $\Phi_l\triangleq\left(\bw^{(l)},\bbb^{(l)}\right)$, are learnt during the training phase.
In 	this work we use a pointwise ReLU activation function $g_l(\cdot)\triangleq \max(0,\cdot)$ yielding the intermediate layer outputs
\setlength{\abovedisplayskip}{10pt}
\setlength{\belowdisplayskip}{10pt}

\[
\by^{(l)}
\triangleq f_l(\bx^{(l)},\Phi_l) =
\begin{cases}
\max(0,\bw^{(l)} \ast \bx^{(l)} + \bbb^{(l)}), & l<L\\
\bw^{(l)} \ast \bx^{(l)} + \bbb^{(l)}, &  l=L
\end{cases}
\]
whose concatenation gives the overall CNN function
\\
\begin{equation}
f(\bx,\Phi) = f_L(f_{L-1}(\ldots f_1(\bx,\Phi_1),\ldots,\Phi_{L-1}),\Phi_L)
\nonumber
\label{eq:chain}
\end{equation}
\\
where $\Phi\triangleq(\Phi_1,\ldots,\Phi_L)$ is the whole set of parameters to learn.

In the proposed solution, the network (Fig. \ref{fig: Net}) is composed by 10 convolutional layers each, except the first and the last, followed by a Rectified Linear Unit (ReLu) activations to ensure fast convergence. The network has a single band image affected by speckle noise $Y$, the overall output is its filtered version 
\setlength{\abovedisplayskip}{20pt}
\setlength{\belowdisplayskip}{20pt}
$$ \hat X=f(\bx,\Phi)$$

\subsection{Training}
The goal of the work is to provide a network for despeckling urban areas. For this aim the CNN is trained on the Google Maps dataset that supply a set of urban images on which speckle is simulated according to (\ref{eq: sar formation}) and (\ref{eq: speckle distribution}). Moreover, in order to give robustness to the network, also a set of generic grayscale images from the UCID and BSD dataset are taking in count for the training.

The training process is performed by the Stochastic Gradient Descent with momentum, with learning rate $ \eta = 2 \cdot 10^{-6}$ on $30000 \times (65 \times 65)$ training patches and $12000 \times (65 \times 65)$ for the validation. 

The cost function $C(\cdot)$ computes the distance between output and reference and according to its value, the parameters $\Phi$ of the network are updated via the SGD optimization process
\setlength{\abovedisplayskip}{10pt}
\setlength{\belowdisplayskip}{10pt}

$$C  = \lambda C_1+ C_2$$

$$C_1 = \textit{SID}( \frac{Y}{\hat{X}},\frac{Y}{X} ) = \textit{SID}(\hat{N},N)$$

$$C_2 =||\hat{X}-X||^2$$

In this work $C(\cdot)$ is a linear combination of two terms: $C_2$  is the mean squared error between filtered image and the noise-free reference; $C_1$ computes a single band adaptation of Spectral  Information Divergence (SID) \cite{Chein99} between the estimated ratio image $\hat N$ and the reference one $N$.
Using $C_2$ ensures to minimize the spatial distance between $\hat X $ and $X$. 
Minimizing $C_1$  makes the network able to predict the speckle noise and preserve its statistical properties.
The aim of using this cost function is two fold: first the network has to predict directly the clean image, second has to take care about the statistical properties of the noise and to do not remove spatial details from the noisy image, but just the speckle.

\begin{table}[t]
	\centering
	\caption{\footnotesize{ \\Hyper-parameters of the proposed network}}
	\small
	\setlength{\tabcolsep}{3pt}
	\begin{tabular}{cccccc}
		\hline
		\multirow{2}{*}{\labsty{Layer}} & \labsty{Features} & \labsty{Kernel}&\labsty{Lerning}&\labsty{Batch} &\multirow{2}{*}{\labsty{ReLU}} \\
		& \labsty{Maps} 	& \labsty{Dimension}&\labsty{Rate} & \labsty{Normalization} & \\
		\hline
		1	& 64 & $3 \times 3$ & $2\cdot 10^{-6}$&False &False \\
		
		2-9	& 64 & $3 \times 3$ & $2\cdot 10^{-6}$&True &True \\
		10	& 1 & $3 \times 3$ & $2\cdot 10^{-6}$&False &False \\
		
		\hline
	\end{tabular}
	\label{tab: architectures}
\end{table}

\section{Experimental results}

In order to assess the performance in an urban environment, the proposed solution is tested on Google Maps images. The networks has never seen these images during the training process. 
In Fig. \ref{fig: output1}-\ref{fig: output2}  is shown a comparison with PPB, one of the most well known solution in the state of art for despeckling.
Although the PPB filtered images seems to be very clean, the proposed solution preserves better the spatial details and give a closer result to the reference. The network seems to remove the noise and to preserve spatial details that in PPB tend to disappear. PPB works well on big scale object like large buildings and roads, but the overall result tends to be over smoothed and so the most of lower scale objects are filtered. The proposed solution is able to generalize the object scale: it can remove the noise saving spatial details at different scales. In fact, cars and trees are still visible in Fig. \ref{fig: output1}, as well as the reconstruction of the roofs in Fig. \ref{fig: output2}. Given that a despeckling solution can be used as pre-processing for other tasks like classification and object detection, preserving objects at different scale plays a very important role in the assessment of performances.\\
Moreover, in Tab. \ref{tab:performance} numerical results are shown. For numerical assessment M-index \cite{Gomez2017} has been computed: this index takes into account the filtering accuracy in both regularizing homogeneous areas, computing the Equivalent Number of Looks (ENL), and preserving structures and details, computing homogeneity of ratio images. An ideal filter would produce an M-index equal to zero.
The values of this index confirm what we say in the visual comparison. 

\begin{table}[t]
	\vspace{0.8cm}
	\caption{\footnotesize \\Numerical Results: M-index evaluated on clip1 and clip2}
	\renewcommand{\arraystretch}{1.2}
	\small
	\centering
	\begin{tabular}{|lcc|}
		\hline
		\labsty{method}	& \labsty{clip1} & \labsty{clip2}\\
		\hline
		Proposed & \textbf{5.59} & \textbf{6.55}\\
		PPB & 10.65& 10.27\\
		\hline
	\end{tabular}
	\label{tab:performance}%
\end{table}%

\begin{figure}
	\begin{tabular}{cc}
		\image{ref1} & \image{noisy1}\\
		\labsty{Reference} & \labsty{Noisy} \\
		\image{ppb1} & \image{prop1}\\
		\labsty{PPB} & \labsty{Proposed} \\
	\end{tabular}
	\caption{Result on simulated data: clip1}
	\label{fig: output1}
\end{figure}

\begin{figure}
	\begin{tabular}{cc}
		\image{ref2} & \image{noisy2}\\
		\labsty{Reference} & \labsty{Noisy} \\
		\image{ppb2} & \image{prop2}\\
		\labsty{PPB} & \labsty{Proposed} \\
	\end{tabular}
	\caption{Result on simulated data: clip2}
	\label{fig: output2}
\end{figure}

\begin{figure*}[t]
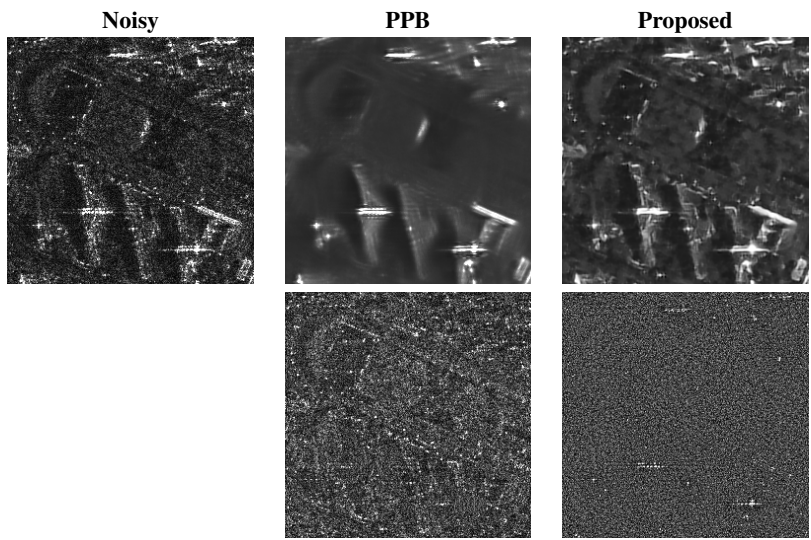

	\centering
	\begin{tabular}{ccc}
		\labsty{Noisy} 		& \labsty{PPB}  			& \labsty{Proposed} \\
		\image{noisyVele} 	& \image{ppbVele} 			& \image{propVele}\\
									& \image{ppbVeleRatio} 	& \image{propVeleRatio}\\
	\end{tabular}
	\caption{Results on real data}
	\label{fig: output_real}
\end{figure*}

\begin{table}[t]
	\vspace{0.8cm}
	\caption{\footnotesize \\Numerical Results: M-index evaluated on real SAR image}
	\renewcommand{\arraystretch}{1.2}
	\small
	\centering
	\begin{tabular}{|lc|}
		\hline
		\labsty{method}	& \labsty{M-index} \\
		\hline
		Proposed & 8.36 \\
		PPB & \textbf{7.29}\\
		\hline
	\end{tabular}
	\label{tab:performance_real}%
\end{table}%

Same considerations can be done for real data: in Fig. \ref{fig: output_real} results on a real SAR images are shown. Without a reference it is difficult to state the quality of a filter, so together with filtered images (top row) we show also the ratio between noisy and filtered image (bottom row). Even if Tab \ref{tab:performance_real} shows a better M-index for PPB, also in this case the proposed solution better preserves details than PPB that tends to present an over-smoothed filtered image as well. Considering the ratio images, it is clear that PPB suppresses a lot of details, meanwhile the proposed solution faces some difficulties filtering strong scatterers.\\

\section{Conclusion and Future Works}

In this work a deep convolutional neural network for despeckling in urban areas is proposed. The network is trained and tested on simulated data. Moreover, the CNN is trained to predict both the clean image and the noise, in order to ensure spatial and statistical consistency in the filtered image. The results are encouraging, the estimated clean images show good details preservation and don't seem to create spatial artefacts on homogeneous areas. In future works, the potential of CNN for despeckling in unsupervised learning will be explored in order to avoid the use of a clean reference.

\bibliographystyle{IEEEbib}
\newcommand*{\bibfont}{\footnotesize}
\bibfont{\bibliography{8_refs_v1}}

\end{document}

%% file: preambol.tex
\usepackage{amsmath,graphicx}
\usepackage{subfig}
\usepackage{color}
\usepackage{epstopdf}
\usepackage{tikz}
\usepackage{balance}
\usepackage{url}
\usepackage{multirow,hhline}
\usepackage{placeins}
\usepackage{wasysym}
\usepackage{amssymb}

\usepackage[nomessages]{fp}
\usepackage{array}

\newcommand{\bx}{\mathbf{x}}
\newcommand{\by}{\mathbf{y}}
\newcommand{\bw}{\mathbf{w}}
\newcommand{\bz}{\mathbf{z}}
\newcommand{\bbb}{\mathbf{b}}

\newcommand{\labsty}[1]{{\small \textbf{#1}}}

\newcommand{\image}{\pgfuseimage}

\newcommand{\figpath}{./Figures/}

\definecolor{colSico}{rgb}{0.2, 0.6, 0.2}

\definecolor{mybegie}{RGB}{128,0,0}

\pgfdeclareimage[width=0.13\textwidth]{ref}{\figpath 02000.jpg}
\pgfdeclareimage[width=0.13\textwidth]{speckle}{\figpath f.png}
\pgfdeclareimage[width=0.13\textwidth]{noisy}{\figpath 02000_1_speckled.png}

\pgfdeclareimage[width=0.23\textwidth]{ref1}{\figpath 01424.jpg}
\pgfdeclareimage[width=0.23\textwidth]{noisy1}{\figpath 01424_noisy.png}
\pgfdeclareimage[width=0.23\textwidth]{ppb1}{\figpath 01424_ppb.png}
\pgfdeclareimage[width=0.23\textwidth]{prop1}{\figpath 01424_cnn.png}

\pgfdeclareimage[width=0.23\textwidth]{ref2}{\figpath 01387.jpg}
\pgfdeclareimage[width=0.23\textwidth]{noisy2}{\figpath 01387_noisy.png}
\pgfdeclareimage[width=0.23\textwidth]{ppb2}{\figpath 01387_ppb.png}
\pgfdeclareimage[width=0.23\textwidth]{prop2}{\figpath 01387_cnn.png}

\pgfdeclareimage[width=0.18\textwidth]{noisyVele}{\figpath vele_noisy.png}
\pgfdeclareimage[width=0.18\textwidth]{ppbVele}{\figpath vele_ppb.png}
\pgfdeclareimage[width=0.18\textwidth]{propVele}{\figpath vele_l2_sid.png}
\pgfdeclareimage[width=0.18\textwidth]{ppbVeleRatio}{\figpath vele_ratio_ppb.png}
\pgfdeclareimage[width=0.18\textwidth]{propVeleRatio}{\figpath vele_ratio_l2_sid.png}

\pgfdeclareimage{net}{\figpath net.pdf}